\documentstyle{article}

\begin{document}
{\large

\begin{center}
\vspace{1cm} {\large \bf Features of Hopping Integral and Unconventional
Orbital Modes for Quantum Skyrmions in Heisenberg Ferromagnet }

\vspace{0.5cm}
{R. A. Istomin and A. S. Moskvin}

\vspace{0.3cm}
{\small
Department of Theoretical Physics, Ural State University,
620083, Lenin Ave. 51, Ekaterinburg , Russia.
}
\vspace{0.5cm}

\end{center}

We use the coherent state approximation for the skyrmion in the isotropic
quantum Heisenberg ferromagnet to obtain analytical expression for skyrmion
hopping integral as a function of distance. It appears  the skyrmion hopping is
restricted by  distances larger than its effective diameter. Some puzzling
properties of the orbital skyrmionic modes are discussed. Spin distribution in
these states is obtained and effect of quantum spin contraction  is
demonstrated.
  The principal possibility of lowering the skyrmion energy due to a bonding-like
  state formation   is illustrated.

PACS numbers 74.60.Ge; 05.40.+j; 30.70;-d; 74.20.-z

\vspace{0.5cm}

\section{Introduction}

Translational motion of the 2D topological excitations, considered as a
particle-like object, is a problem  of enhanced importance both in theoretical
and practical sense. Classical approach implies making use of phenomenological
equations of motion. A variational approach is one of the most attractive
methods to study an appropriate quantum problem. The variational wave function
of collective motions, like translations, rotations and oscillations, was
initially proposed by Hill and Wheeler $\cite{Hill}$ to be
\begin{equation}
\Psi ([{\bf r}])=\int d^{2}{\bf R} f({\bf R})\Psi _{{\bf R}} ([{\bf r}]),
\end{equation}
where $\Psi _{{\bf R}} ([{\bf r}])$ is a particle-like wave function for the
topological excitation centered at point ${\bf R}=(R,\Theta)$, $f({\bf R})$ a
weighting function, and $[{\bf r}]$ is a set of particle coordinates.  For
continuous 2D media the angular dependence of the weighting function follows
immediately from the cylindrical symmetry of the system, so,
$$
f({\bf R})=f_{l}(R)\exp(il\Theta)
$$
 with $l$ being the angular momentum. By
varying the weighting function to minimize the expectation value of the
Hamiltonian
$$
E=\langle \Psi ([{\bf r}])|{\hat H}|\Psi ([{\bf r}])\rangle,
$$
one  obtains an integral equation for $f({\bf R})$,
\begin{equation} \label{v}
\int d^{2}{\bf R}' K({\bf R}',{\bf R}) f({\bf R}')=0,
\end{equation}
where the kernel is
\begin{equation} \label{vv}
K({\bf R}',{\bf R})=\langle \Psi _{{\bf R}'} ([{\bf r}])|{\hat H}-E|\Psi _{{\bf
R}}([{\bf r}])\rangle=H({\bf R}',{\bf R})-ES({\bf R}',{\bf R}),
\end{equation}
where we have made use of matrix elements of Hamiltonian, and overlap integral
$S({\bf R}',{\bf R})$. The kernel $K({\bf R}',{\bf R})$ may be considered as an
effective transfer (hopping) integral for quasiparticle transfer between ${\bf
R}'$ and $ {\bf R}$ sites.

It should be noted however that full realization of above mentioned scheme (\ref{v},\ref{vv}) for the real topological defect on the lattice is a difficult task as the lattice has no authentic rotational symmetry unlikely the conventional case of the vortex in the superfluid \cite{Tang}. Obviously, the first step to treat this problem implies using continual approximation for describing soliton in the lattice.

 Below we calculate quantity $K({\bf R}',{\bf R})$ for quantized  skyrmion being peculiar topological defect described in the framework of  the coherent state approximation and use continual approximation to treat skyrmion motion in a discrete lattice. 

Skyrmions are general static solutions of  2D Heisenberg ferromagnet with
isotropic  spin-Hamiltonian
\begin{equation}
H=-J\Sigma_{i,b}{\bf S}_{i}{\bf S}_{i+b}, \label{H}
\end{equation}
 obtained by Belavin and Polyakov $\cite{BP}$ from classical nonlinear sigma model. A renewed interest
to these unconventional spin textures is stimulated by high-$T_c$ problem in
doped quasi-2D-cuprates and  quantum Hall effect.

The skyrmion spin texture consists of a vortex-like arrangement of the in-plane
components of spin with the $z$-component reversed in the centre of the
skyrmion and gradually increasing to match the homogeneous background at
infinity. The spin distribution within classical skyrmion  is given as follows
\begin{equation}
S_{x}=\frac{2r\lambda }{r^{2}+\lambda ^{2}}\cos \varphi ,\qquad
S_{y}=\frac{2r\lambda }{r^{2}+\lambda ^{2}}\sin \varphi ,\qquad
S_{z}=\frac{r^{2}-\lambda ^{2}}{r^{2}+\lambda ^{2}}.
\end{equation}
In terms of the stereographic variables the skyrmion with radius  $\lambda $   and
phase  $\varphi _{0}$  centered at a point $z_0$ is identified with spin distribution
 $w(z)=\frac{\Lambda }{z-z_0}$, where $z=x+iy=re^{i \varphi }$  is a point in the complex
 plane, $\Lambda =\lambda e^{i\theta}$,
and characterized by three modes: translational, or positional $z_0$-mode,
"rotational" $\theta$-mode and    "dilatational" $\lambda$-mode. Each of them
corresponds to certain symmetry of the classical skyrmion configuration. For
example, $\theta$-mode corresponds to combination of rotational symmetry and
internal $U(1)$ transformation.

The simplest wave function of the quantum spin system, which corresponds to
classical skyrmion, is a product of spin coherent states $\cite{Perelomov}$. In
case of spin $s=\frac{1}{2}$
\begin{equation}
\Psi _{sk}( 0) =\prod\limits_{i}\lbrack \cos \frac{\theta
_{i}}{2}e^{i\frac{\varphi _{i}}{2}}\mid \uparrow \rangle +\sin \frac{\theta
_{i}}{2}e^{-i\frac{\varphi _{i}}{2}}\mid \downarrow \rangle \rbrack ,
\label{CS}
\end{equation}
where $\theta _{i}=\arccos \frac{r_{i}^{2}-\lambda ^{2}}{r_{i}^{2}+\lambda ^{2}}$. Coherent state implies a maximal equivalence to classical state with minimal uncertainty of spin components.

Classical skyrmions with different phases and radia have equal energy.
Nevertheless, stationary state of quantum skyrmion is not a superposition of
states with different phase and radius $\cite{Stern}$ and has a certain
distinct value of $\lambda$. Fluctuations in the in-plane orientation and
 size $\lambda$ for quantum skyrmion are intimately connected; rotating a skyrmion changes
its size. This result follows from the commutation relations of quantum angular
momentum operators \cite{Green}.

In our recent paper $\cite{Istomin}$ it has been obtained an analytical
expression for overlap integral of skyrmion states
\begin {equation}
S({\bf R}_{1},{\bf R}_{2})=\exp[-\frac{\pi S}{a^{2}}(R_{12}^{2}-4i\Delta
_{12})], \label{S}
\end {equation}
if $R_{12}< 2\lambda$
or
$$
S({\bf R}_{1},{\bf R}_{2})=\exp[-\frac{\pi S}{a^{2}}( R_{12}^{2}-4i\Delta _{12}
$$
\begin {equation}-R_{12}\sqrt{R_{12}^{2}-
4 \lambda^{2}}+2 \lambda^{2}ln(\frac{R_{12}+\sqrt{R_{12}^{2}-4
\lambda^{2}}}{R_{12}-\sqrt{R_{12}^{2}-4 \lambda^{2}}})],
 \label{SS}
\end {equation}
if $R_{12}> 2\lambda $. Here one introduces a quantity
$$
\Delta _{12}=\frac{1}{2} \mid R_{1}\mid \mid R_{2}\mid \sin( \varphi _{1}-\varphi _{2})
$$
being area of a sector skyrmion covers while moving in the plane, $a$ is a
lattice parameter.
 Skyrmion, having wandered closed contour on the plane,
acquires the phase $4\pi\frac{\Delta _{12}}{a^{2}}$, that is  each spin lying
inside the contour
 contributes  $4\pi$ into the total phase of the skyrmion, so its
motion appears to be  likely the motion of a charged particle in magnetic field
$\cite{Stone,NS,Volovik}$. Extending this analogy one may introduce the effective
magnetic length $d=(\frac{4 \pi S}{a^{2}})^{-\frac{1}{2}}$.
 Note the dependence of the phase factor in the overlap
integral on the origin of the vectors ${\bf R}_1$, ${\bf R}_2$. It should be
noted that for large separation $R_{12}\gg \lambda$ the $R$ dependence of the
overlap integral obeys the power law
$$
S({\bf R}_{1},{\bf R}_{2})\propto (\lambda /R_{12})^p
$$
with $p=(\lambda /d)^2$.

\section{Hopping integral for stiff quantum skyrmion}
One of the main goals of the paper is a calculation of the nondiagonal matrix
element of the Hamiltonian (\ref{H})
$$
H({\bf R}_{1},{\bf R}_{2})= \langle \Psi ^{sk}_{{\bf R}_{1}} ([{\bf r}])|{\hat
H}|\Psi ^{sk}_{{\bf R}_{2}}([{\bf r}])\rangle ,
$$
 between the states of quantum  skyrmions,
located at points $R_{1}$ and  $R_2$, respectively, or transfer integral,
likely being the principal quantity defining the quasiparticle motion of this
topological defect.

We use approximation of  "stiff" skyrmions, which takes no  account of internal
skyrmion quantum fluctuations and is closely related to classical limit. Making
use of the coherent states (\ref{CS}) we write out the appropriate matrix
element as follows:
\begin{equation}
H({\bf R}_{1},{\bf R}_{2})= -\frac{1}{2}JS_{12}\Sigma_{i,b}
\frac{\langle\varsigma_{i}\mid {\bf S}_{i}
\mid\varsigma_{i+b}\rangle\langle\varsigma_{i+R_{12}}\mid {\bf S}_{i+R_{12}}
\mid\varsigma_{i+b+R_{12}}\rangle}{\langle\varsigma_{i}
 \mid\varsigma_{i+b}\rangle\langle\varsigma_{i+R_{12}}\mid
\varsigma_{i+b+R_{12}}\rangle}.
\end{equation}

For the spin coherent states the following relations exist $\cite{Radcliffe}$:
$$
\langle\varsigma\mid
S_{z} \mid\mu\rangle=-S\frac{( 1-\varsigma \overline{\mu })( 1+\varsigma \overline{\mu }) ^{2S-1}}{( 1+\mid \varsigma \mid ^{2}) ^{S}( 1+\mid \mu \mid ^{2}) ^{S}},
$$
$$
\langle\varsigma\mid
S_{+} \mid\mu\rangle=2S\frac{\bar{\mu} ( 1+\varsigma \overline{\mu }) ^{2S-1}}{( 1+\mid \varsigma \mid ^{2}) ^{S}( 1+\mid \mu \mid ^{2}) ^{S}},
$$
\begin{equation}
\frac{\langle\varsigma\mid
S_{z} \mid\mu\rangle}{\langle\varsigma
 \mid\mu\rangle}=-S\frac{ 1-\varsigma \overline{\mu} }{ 1+\varsigma \overline{\mu} },
 \label{Sz}
\end{equation}
$$
\frac{\langle\varsigma\mid
S_{+} \mid\mu\rangle}{\langle\varsigma
 \mid\mu\rangle}=2S\frac{ \overline{\mu}}{ 1+\varsigma \overline{\mu} },
\quad \frac{\langle\varsigma\mid S_{-} \mid\mu\rangle}{\langle\varsigma
 \mid\mu\rangle}=2S\frac{ \varsigma }{ 1+\varsigma \overline{\mu} }.
$$
Taking account of the spin distribution in the skyrmion we write out the matrix
element addressed as follows
$$
H({\bf R}_{1},{\bf
R}_{2})=-\frac{1}{2}JS^{2}S_{12}\Sigma_{i,b}\lbrack\lbrack((\varsigma_{i}-R_{1})
(\overline{\varsigma_{i}}-\overline{R_{2}})-\lambda^{2})((\varsigma_{i+b}-R_{1})\times
$$$$\times(\overline{\varsigma_{i+b}}-\overline{R_{2}})-\lambda^{2})+
2\lambda^{2}(\varsigma_{i+b}-R_{1})(\overline{\varsigma_{i}}-\overline{R_{2}})
+2\lambda^{2}(\overline{\varsigma_{i+b}}-\overline{R_{2}})(\varsigma_{i}-R_{1})\rbrack
$$
\begin{equation}
\lbrack(\lambda^{2}+(\varsigma_{i}-R_{1})
(\overline{\varsigma_{i}}-\overline{R_{2}}))(\lambda^{2}+(\varsigma_{i+b}-R_{1})
(\overline{\varsigma_{i+b}}-\overline{R_{2}}))\rbrack^{-1}\rbrack . \label{ME}
\end{equation}
Below we are going to find the limit  of the right hand side when $a\rightarrow
0$ with the constant $a$ of lattice assumed to be square. To do it one has to
make expansion in lattice constant $a$ and to keep the three first terms,
 then to turn the sum into integral. The first (zero order in $a$) term in the
expansion, as it is easy to see, is simply the ground state energy. The linear
in $a$ term in the expansion is zero due to the inversion symmetry. So, one has
to  define the second order term for a given site $i$.
Straightforward calculation gives for that second order term in  $a$ in the matrix element for scalar
product $\Sigma_{b}{\bf S}_{i}{\bf S}_{i+b}$ the following result
$$
\frac{-8\lambda^{2}}{(\lambda^{2}+
(\overline{\varsigma}-\overline{R_{2}})(\varsigma-R_{1}))^{2}}a^{2}.
$$
While coming to the limit $a\rightarrow 0$  the summation in ($\ref{ME}$) is
replaced by the integration, and this term results in a nonvanishing
contribution to the transfer integral:
$$
H({\bf R}_{1},{\bf R}_{2})=\lbrack
E_{0}+4JS^{2}\lambda^{2}\int^{\infty}_{0}rdr
$$
\begin{equation}
\int^{2\pi}_{0}d\varphi \frac{1}{(\lambda^{2}+
(re^{-i\varphi}-\overline{R_{2}})(re^{i\varphi}-R_{1}))^{2}}\rbrack S({\bf
R}_{1},{\bf R}_{2}), \label{HRR}
\end{equation}
 where $E_{0}$ is a ground state energy of spin system. An angular integral  is given
 by following expression
$$\int^{2\pi}_{0}d\varphi
\frac{1}{(\lambda^{2}+
(re^{-i\varphi}-\overline{R_{2}})(re^{i\varphi}-R_{1}))^{2}}=$$
 $$
\int \frac{zdz}{i\,(z-x_{1})^{2}(z-x_{2})^{2}(r\overline{R_{2}})^{2}}=
\frac{2\pi(x_{1}+x_{2})}{(r\overline{R_{2}})^{2}(x_{1}-x_{2})^{3}} ,
$$
so the radial integral is
$$
\int^{\infty}_{0} \frac{ 2\pi
r(r^{2}+R_{1}\overline{R_{2}}+\lambda^{2})dr}{\sqrt{((r^{2}+R_{1}\overline{R_{2}}+\lambda^{2})^{2}-4r^{2}R_{1}\overline{R_{2}})^{3}}}=\frac{1}{\lambda^{2}},
$$
where $$
x_{1,2}(r,R_{1},R_{2})=\frac{\lambda^{2}+r^{2}+R_{1}\overline{R_{2}}\pm K}{2r\overline{R_{2}}},
$$
$$
K(r,R_{1},R_{2})=\sqrt{(\lambda^{2}+r^{2}+R_{1}\overline{R_{2}})^{2}-4r^2R_{1}\overline{R_{2}}},
$$
where the complex root has positive real part.

When  $R_{12}<2\lambda$ we finally obtain for the transfer integral
\begin{equation}
H({\bf R}_{1},{\bf R}_{2})=\lbrack E_{0}+4\pi JS^{2}\rbrack S({\bf R}_{1},{\bf
R}_{2}). \label{transfer}
\end{equation}
  As $R_{12}=0$ $H({\bf R}_{1},{\bf R}_{1})=\lbrack
E_{0}+4\pi JS^{2}\rbrack,$ and this is the classical skyrmion energy. The
structure of the transfer integral $H({\bf R}_{1},{\bf R}_{2})$ is such, that
its contribution to
 the effective hopping integral (3) $K({\bf R}_{1},{\bf R}_{2})$ at $R_{12}<2\lambda$
 is exactly compensated by the overlap
contribution. Thus, effective hopping integral for the "stiff" skyrmion is zero
for distances $R_{12}<2\lambda$.

When  $R_{12}>2\lambda$, the integral ($\ref{HRR}$) contains two singular points
and, respectively, it is ill-defined in that region. To avoid the contribution
from these points and obtain result independent of choice of coordinates we
calculate its principal value,
 that is  remove the contribution of points, that lie inside the circles of small radius
 $\epsilon$  with centers in singular points. To this end we divide the plane
 into two equal parts drawing the border between halfplanes at the middle of
 the line connecting two singular points. The value of integral ($\ref{HRR}$)
  is twice the value of integral over halfplane.
Integral ($\ref{HRR}$) being represented as halfplane integral  in coordinate
system with center in  singular point   takes the form
$$
H({\bf R}_{1},{\bf R}_{2})=\lbrack E_{0}+8JS^{2}\lambda^{2}\int_{c}^{\infty
}rdr \times
$$
\begin{equation}
\times \int^{2\pi-\arccos{\frac{c}{r}}}_{\arccos{\frac{c}{r}}}d\varphi
\frac{1}{r^{2}(r+
(\frac{R_{12}}{2}-c)e^{-i\varphi}-(\frac{R_{12}}{2}+c)e^{i\varphi})^{2}}+
\label{HRRR}
\end{equation}
$$+8JS^{2}\lambda^{2}\int ^{c}_{0}rdr\int^{2\pi}_{0}d\varphi
\frac{1}{r^{2}(r+
(\frac{R_{12}}{2}-c)e^{-i\varphi}-(\frac{R_{12}}{2}+c)e^{i\varphi})^{2}}\rbrack
S({\bf R}_{1},{\bf R}_{2}),
$$
where $c=\sqrt{\frac{R_{12}^{2}}{4}-\lambda^{2}}$ is a half-separation between
 singular points. Here we have restricted ourselves to the case of real ${\bf R}_{1},{\bf R}_{2}$
to simplify calculations. It is easy to see that last term in ($\ref{HRRR}$) is zero as angular
integral is zero no matter what the value of $r$ is. Then,
\begin{equation}
 \int d\varphi \frac{1}{(r+
(\frac{R_{12}}{2}-c)e^{-i\varphi}-(\frac{R_{12}}{2}+c)e^{i\varphi})^{2}}=
\end{equation}
$$
\frac{1}{i(r^{2}+4\lambda^{2})}[-\frac{e^{i\varphi}}{e^{i\varphi}-x_{-}}-\frac{e^{i\varphi}}{e^{i\varphi}-x_{+}}
-\frac{x_{-}+x_{+}}{x_{-}-x_{+}}\ln{\frac{e^{i\varphi}-x_{-}}{e^{i\varphi}-x_{+}}}],
$$
where
$$
x_{+,-}=\frac{r\pm\sqrt{r^2+4\lambda^{2}}}{2(\frac{R_{12}}{2}+c)}.
$$
 As a result, we obtain the following radial integral
$$
\int
_{c}^{\infty}rdr\int^{2\pi-\arccos{\frac{c}{r}}}_{\arccos{\frac{c}{r}}}d\varphi
\frac{1}{r^{2}(r+
(\frac{R_{12}}{2}-c)e^{-i\varphi}-(\frac{R_{12}}{2}+c)e^{i\varphi})^{2}}=
$$
\begin{equation}
\int^{\infty}_{c}dr\frac{2}{r(r^{2}+4\lambda^{2})}[\frac{r}{\sqrt{(r^{2}+4\lambda^{2})}}\times
\end{equation}
$$
\times\arctan{\sqrt{(r^{2}+4\lambda^{2})(\frac{1}{c^2}-\frac{1}{r^{2}})}}-
\frac{\sqrt{r^{2}-c^{2}}(2\lambda^{2}+r^{2})2c}{-4c^{2}\lambda^{2}+r^{2}(r^{2}+4\lambda^{2})}]=0.
$$

Thus, it appears  that the effective hopping integral is nonzero in this regime
and its value is simply proportional to the overlap integral.

 So,  the final result for effective hopping integral is
\begin{equation}
  K({\bf R}_{1},{\bf R}_{2})=-4\pi S^{2}J \,S({\bf R}_{1},{\bf R}_{2})\theta(R_{12}-2\lambda),
\end{equation}
where
$$
 \theta(R_{12}-2\lambda)=\cases{1,& if $R_{12}>2\lambda$ \cr 0,&
otherwise \cr},
$$
and $S({\bf R}_{1},{\bf R}_{2})$ is given by  expressions ($\ref{S}$),
($\ref{SS}$).

Thus, one might conclude that the skyrmion hopping is restricted by  distances
larger than $ 2\lambda $. The nonzero effective hopping integral for distances
larger than $ 2\lambda $ allows to lower skyrmion energy by construction of
appropriate combination of skyrmion states centered at different points of
lattice and having definite value of angular moment. Energy of these states
will depend on the value of the angular moment and on number of the levels as
it is the case for the Landau levels in a magnetic field. Note also that
 transfer integral has the negative sign as must be the case.

The states considered have definite phase and radius. In fact phase and radius
can not be treated separately as radius is determined by spin of the skyrmion
and spin in turn is canonically conjugate to the phase. It is easy to show that
states of definite phase and radius are not mixed by the Hamiltonian. This is
related to the fact that we deal with the skyrmion of charge $q=1$, which has
infinite spin. As the spin and phase operators do not commute, the spin must
have infinite value for the phase to be definite. In the case of higher
topological charges there would be mixing of the states with different phases
and radius. The similar conclusions were drawn by Stern $\cite {Stern}$.

 \section{Orbital motion of skyrmions. Small and large orbital $l$-skyrmions}

 Making use of the above derived results we may exploit particular analogy between
skyrmion and particle in zero Landau level to construct a subset of orthogonal
states with definite value of orbital momentum, or $l$-skyrmions. Consider
firstly the following orbital skyrmionic modes being linear combinations of
single skyrmion states
\begin {equation}\Psi ^{sk}_{l}(R)=\int_{0}^{2\pi}\exp{(il\Theta)}\Psi
^{sk}_{R\exp{(i\Theta)}}d\Theta, \label{Psi}
\end{equation}
where $\Psi ^{sk}_{R\exp{(i\Theta)}}$ is a single skyrmion state centered at
the point $Z=R\exp{(i\Theta)}.$ Below, we will distinguish small and large
$l$-skyrmions depending on whether $R<\lambda$ or $R>\lambda$, respectively.

\subsection{Small  orbital $l$-skyrmions}
It seems, any two equally centered $\Psi ^{sk}_{l}(R_{1})$ and $\Psi
^{sk}_{l}(R_{2})$ states with commonly different $R_{1}<\lambda$ and
$R_{2}<\lambda$, but the same orbital momentum $l$ are linearly dependent.
Indeed, for small $l$-skyrmions one might use a simple  Gaussian form ($\ref{S}$)
for the overlap integral of single skyrmionic states, so the corresponding
unnormalized overlap integral is given by a rather simple analytical expression
$$
S(R_{1},R_{2})=\int_{0}^{2\pi}\exp[il(\Theta_{1}-\Theta_{2})]\exp[-\frac{1}{4d^{2}}(R_{1}^{2}+R_{2}^{2}-2
R_{1}R_{2}
$$
\begin{equation}
\exp[-i(\Theta_{1}-\Theta_{2})])]d\Theta_{1}d\Theta_{2}
 =\frac{4\pi^{2}}{l!}\exp[-\frac{1}{4d^{2}}(R_{1}^{2}+R_{2}^{2})](\frac{R_{1}R_{2}}{2d^{2}})^{l}.
\end{equation}
  It is easy to see
 that after normalization the overlap integral for two $l$-skyrmions with
 different $R_1$ and $R_2$
 $$
 \frac{S(R_{1},R_{2})}{\sqrt{S(R_{1},R_{1})S(R_{2},R_{2})}}=1
 $$
 does not depend on $R_{1},R_{2}$ at all, which means at least that the corresponding functions
  are  not independent. So, these states
are distinguished by a single quantum number $l$ which is an orbital momentum
and are degenerate in energy in full analogy with a zero Landau level. It
should be emphasized that for "$l$-skyrmion" $\Psi ^{sk}_{l}(R)$ with
$R<\lambda$ the orbital moment is restricted by non-negative values $l\geq 0$,
due to the definite sign of the Berry phase in the overlap integral for single
skyrmion. To describe a spatial distribution of a "single skyrmionic
(quasiparticle) density" for the orbital $l$-skyrmion $\Psi _l$ one might use
an overlap integral of the orbital state ($\ref{Psi}$) with the simple skyrmion
state in the point $z=r\exp{i\theta}$ which has a rather simple form for
$r<\lambda$
$$
S^{l}_{sk}(r,R)=\int_{0}^{2\pi}\exp[il\Theta
]\exp[-\frac{1}{4d^{2}}(r^{2}+R^{2}-2 rR\exp[-i(\theta -\Theta )])]d\Theta =
$$
 \begin{equation}
 \frac{2\pi}{l!}\exp[il\theta
 ]\exp[-\frac{1}{4d^{2}}(r^{2}+R^{2})](\frac{rR}{2d^{2}})^{l},
\end{equation}
 that with account for normalization conditions gives the quantity
\begin{equation}
S^{l}_{sk}(r)=\frac{S_{sk}(r,R)}{\sqrt{S(R,R)}}=\frac{1}{\sqrt{l!}}
\exp[il\theta ](\frac{r^{2}}{2d^{2}})^{l}\exp[-\frac{r^{2}}{2d^{2}}],
\end{equation}
which does not depend at all  on $R$ value. This distribution function has
maximum at $r=r_l$, where
\begin{equation}
\frac{2\pi S r_{l}^{2}}{a^{2}}=l, \label{r}
\end{equation}
so, the orbital momentum $l$ defines the characteristic "orbital area" $\pi
r_{l}^{2}$ measured in terms of a "flux quantum" $\frac{a^{2}}{2 S}$. This
relation looks like a specific quantization condition for the "orbital area".
The $r_l$ value defined by expression ($\ref{r}$) may be called as a
characteristic "internal" radius of the $\Psi_{l}(R)$ orbital states which
magnitude does not depend on $R$ value. One should note that for "s-skyrmion"
$r_0 =0$ as it should be for typical quasiparticle orbital s-states.

Actually, the maximal value of orbital momentum for the small $l$-skyrmion
  is restricted by the size of the bare
skyrmion and, in general, can not significantly exceed $l_{\lambda }$, because of the
number of translational degrees of freedom  is restricted by the "skyrmionic
area" $\pi\lambda^{2}$. If we consider values of $l$ much larger than $l_{\lambda }$
we would obtain meaningless result for the spin distribution in the $l$-skyrmion 
because in this case continual approximation for the skyrmion overlap breakes down and 
expression (\ref{S}) that neglects discretness effects is not adequate.

\subsection{Quantum spin contraction for small radius $l$-skyrmions}
 Spin distribution in
the puzzling "$l$-skyrmionic" $\Psi_{l}(R)$ states with small $R<\lambda$ can
also be shown to be independent on  radius $R$. Besides, the corresponding spin
texture is characterized by a quantum spin contraction, that is the  mean spin
length is less than $S$. It can be calculated with making use of relations
($\ref{Sz}$), so that we have for the spin averages
$$
S_{+}(z)=\frac{1}{N}\int_{0}^{2\pi}\exp{[il(\theta_{1}-\theta_{2})]}\frac{\langle
\frac{(z-z_{1})}{\lambda}\mid S_{+}
\mid\frac{(z-z_{2})}{\lambda}\rangle}{\langle\frac{(z-z_{1})}{\lambda}
 \mid \frac{(z-z_{2})}{\lambda}\rangle}S_{12}(z_{1}-z_{2})d\theta_{1}d\theta_{2}=
$$
$$
=\frac{1}{N}\int_{0}^{2\pi}\exp{[il(\theta_{1}-\theta_{2})]}S\frac{ 2
\lambda(z-r\exp{(i\theta_{1})})
}{\lambda^{2}+(z-r\exp{(i\theta_{1})})(\bar{z}-r\exp{(-i\theta_{2})})
}\times
$$
\begin{equation}
\times \exp[-\frac{1
}{2d^{2}}(r^{2}-r^{2}\exp{[-i(\theta_{1}-\theta_{2})]})]d\theta_{1}d\theta_{2},
\end{equation}
where $z_{1}=r\exp{i\theta_{1}},z_{2}=r\exp{i\theta_{2}}$ are locations of
distinct skyrmion states on the circle of radius $r$,
$$
N=S(r,r)= \frac{(2 \pi)^2}{l!}(\frac{r^{2}}{2d^{2}})^{l}
\exp[-\frac{r^{2}}{2d^{2}}]
$$
is a normalization integral. Introducing complex variables
$\acute{z}_{1}=\exp(i\theta_{1}),\acute{z}_{2}=\exp(-i\theta_{2})$ and taking
integral in $\acute{z}_{1}$, and then in $\acute{z}_{2}$ we obtain
\begin{equation}
S_{+}(z)=\frac{1}{N}\int\int\acute{z}_{1}^{-l+1}\acute{z}_{2}^{-l+1}\frac{2
\lambda(z-r\acute{z}_{1})}
{\lambda^{2}+(z-r\acute{z}_{1})(\bar{z}-r\acute{z}_{2})} \exp[-\frac{1
}{2d^{2}}(r^{2}-r^{2}\acute{z}_{1}\acute{z}_{2})]d\acute{z}_{1}d\acute{z}_{2}
\label{S+}
\end{equation}
$$=\frac{1}{l!}(\frac{2d^{2}}{r^{2}})^{l}\lim_{\acute{z}_{2}\rightarrow 0}\frac{d^{l}}{d\acute{z}_{2}^l}
\lim_{\acute{z}_{1}\rightarrow 0}\frac{d^{l}}{d\acute{z}_{1}^l}[\frac{2
\lambda(z-r\acute{z}_{1})}
{\lambda^{2}+(z-r\acute{z}_{1})(\bar{z}-r\acute{z}_{2})}
\exp[\frac{r^{2}}{2d^{2}}\acute{z}_{1}\acute{z}_{2}]].
$$
This is an exact expression as integrals in ($\ref{S+}$) have no other poles but
those of $(l+1)$-order in $\acute{z}_{1}=0$ and $\acute{z}_{2}=0$, as long as
$r<l.$ For the $S_{z}$ we have a similar expression

\begin{equation}S_{z}(z)=\frac{1}{l!}(\frac{2d^{2}}{r^{2}})^{l}\lim_{\acute{z}_{2}\rightarrow 0}\frac{d^{l}}{d\acute{z}_{2}^l}
\lim_{\acute{z}_{1}\rightarrow
0}\frac{d^{l}}{d\acute{z}_{1}^l}[\frac{\lambda^{2}-(z-r\acute{z}_{1})(\bar{z}-r\acute{z}_{2})}
{\lambda^{2}+(z-r\acute{z}_{1})(\bar{z}-r\acute{z}_{2})} \exp[\frac{1
}{2d^{2}}r^{2}\acute{z}_{1}\acute{z}_{2}]].\label{sz}
\end{equation}
 Surprisingly, for the case $l=0$ we obtain the same
spin distribution as in the simple single skyrmion state. In other words, in
the framework of a classical approach the s-skyrmion is equivalent to a single
skyrmion. For the "p-skyrmion" with $l=1$ the straightforward calculation gives
$$
S_{+}(z)= \frac{2\lambda
z}{\lambda^{2}+z\bar{z}}-\frac{8\lambda d^{2}z}{(\lambda^{2}+z\bar{z})^{2}}+
\frac{8\lambda d^{2} z^{2} \bar{z}}{(\lambda^{2}+z\bar{z})^{3}} =
\frac{2\lambda z}{\lambda^{2}+z\bar{z}}[1-\frac{4d^{2}
\lambda^2}{(\lambda^{2}+z\bar{z})^{2}}]
$$
\begin{equation}
S_{z}(z)=\frac{\lambda^{2}-z\bar{z}} {\lambda^{2}+z\bar{z}}[1-\frac{4d^{2}
\lambda^2}{(\lambda^{2}+z\bar{z})^{2}}], \label{S++}
\end{equation}
  where all values are
measured in lattice constant.  Here, we deal with a spin length contraction,
compared with the case of  classical skyrmion, specified by an universal
function of skyrmion radius and $r=\mid z \mid$
$$
F(\lambda ,r)=[1-\frac{4d^{2} \lambda^2}{(\lambda^{2}+z\bar{z})^{2}}].
$$
 Increasing
the skyrmion radius entails decreasing the spin contraction. Minimal spin
contraction takes place for large $r$'s and $\lambda$'s, while the maximal spin
contraction does in the center of the spin texture for small $\lambda$'s.
Interestingly, that for small skyrmion with $\lambda = 2d$ the effective spin
length in the core of the orbital $p$-skyrmion turns to zero: $S(r=0)=0$.
However, as it is easy to see for very small bare skyrmions with $\lambda \ll
d$ the expression ($\ref{S++}$) leads to unphysical results. Indeed, the
continuous approximation used above appears to be irrelevant for small
skyrmions with radius $\lambda $ less than the effective magnetic length $d$.

 The case $l>1$ can not be represented
in the simple form like ($\ref{S++}$), however, with increase of $l$ the value of
spin contraction in the center of spin distribution also increases. It should
be noted, that  for the orbital skyrmions with  $l=l_{\lambda
}=\frac{2\pi\lambda^{2}S}{a^{2}}=\frac{\lambda ^{2}}{2d^{2}}$ (which is just  a
ratio of "skyrmionic area" $\pi\lambda^{2}$ to the area corresponding to the
quantum of flux $\frac{a^{2}}{2 S}$), the mean spin length in the center of
spin distribution turns to zero, that is the spin contraction appears to be
maximal. This is due to the fact that at $l=l_{\lambda }$ the characteristic
internal radius $r_{l=l_{\lambda }}$ of the $\Psi_{l_{\lambda }}$ orbital
skyrmion defined by expression ($\ref{r}$) coincides with skyrmion radius
$\lambda$, so in the $l_{\lambda }$-skyrmion core we deal with a superposition
of vectors with zero $z$-projection and all possible phases.

\subsection{Large radius orbital skyrmions}
Above we addressed to the particular type of the small orbital $l$-skyrmion
with bare radius $R<\lambda$. Unfortunately, for large $l$ strongly exceeding $l_{\lambda}$ continual approximation breakes down so we can not use expressions (\ref{S+},\ref{sz}) to calculate spin distribution
in the $l$-skyrmions with large effective radius. But this problem may be solved by taking 
bare radius $R$ in the $l$-skyrmion equal to corresponding effective radius 
$r_{l}=\frac{la^{2}}{2\pi S}$. However, in this case we cannot make use of advantages provided by a
simple  Gaussian form for the overlap integral of single skyrmionic states
available only for $R_{12}<2\lambda$ (see ($\ref{S}$)). The orbital motion with large
bare radius acquires some unconventional properties compared to the case of
"small orbits". Spin distribution in these states can be well approximated by the following expressions
\begin{equation}\label{RSz}
S_{z}(z)=-1+\frac{2 \lambda^{2}}{\sqrt{(\lambda^{2}+z\bar{z}+R^{2})^{2}-4 z\bar{z}R^{2}}},
\end{equation}
\begin{equation}
S_{+}(z)=\frac{\lambda}{\bar{z}}(1+\frac{z\bar{z}-\lambda^{2}-R^{2}}{\sqrt{(\lambda^{2}+z\bar{z}+R^{2})^{2}-4 z\bar{z}R^{2}}}).\label{RS+}
\end{equation}
This expressions can be easily obtained if one neglects overlap of skyrmions in different points at all.
Here, R equals to $r_{l}=\frac{la^{2}}{2\pi S}$ as was explained before. Numerical calculations show that
if we take account of the skyrmion overlap the effective radius of the state with bare radius 
$R$ may be changed but orbital momentum can not strongly deviate from $\frac{2\pi SR^{2}}{a^{2}}$ because otherwise discretness effects become again significant.  Expression (\ref{RSz}) shows that maximal 
spin contraction occurs not in the origin as is the case for small skyrmions but close to the points with $\mid z\mid=R.$ 

Strictly speaking, the correct description of large skyrmion implies that one has keep both
bare radius $R$ and orbital momentum $l$ to characterize it because matrix 
element of the Hamiltonian and overlap integral are nontrivial 
when $R_{12}>\lambda$. But for large skyrmion radia such a description is  
redundant at least when we are interested in the properties of the zero 
Landau level only. The corresponding effects would be negligible.

As to the energy of these states its difference from the classical skyrmion 
energy is negligibe for large skyrmion radia so both large and small skyrmions
 may be said  to constitute zero Landau level of skyrmion motion. In principle, 
   it is possible to obtain higher Landau levels but we think it requres 
   detailed
  account of discretness effects and is beyond the scope of this paper. 
  The calculation of the fine structure of the skyrmion energy due to 
  translational effects is  prevented by discretness effects in the framework of the present 
  continual approximation since it  does not allow to have arbitrary values of 
  bare radius $R$ for a given value of $l$ so one can
not realize variational procedure (\ref{v},\ref{vv}) similarly to the case of 
a vortex in superfluid \cite{Tang}.  

 Both types of orbital
skyrmions represent very interesting objects for further investigations.

\section{Conclusions}
Making use of the coherent state approximation allows to obtain an analytical
form for an effective hopping integral for single stiff Belavin-Polyakov
skyrmion in 2D isotropic ferromagnet. It appears,  the skyrmion hopping is
restricted by  distances larger than its effective diameter, thus resulting in
a very unusual quasiparticle properties. We introduce two types of orbital
skyrmionic modes, the small and large $l$-skyrmions, respectively. For the
former it is found both the spatial distribution of a single skyrmionic
density, and the mean spin density distribution. The small $l$-skyrmions with
nonzero orbital momentum  are characterized by a  strong quantum spin
contraction effect.
  Translational symmetry of the Heisenberg Hamiltonian and topological Berry phase in overlap
integral implies that there exists highly degenerate set of skyrmion states in
analogy with a zero Landau level. Based on the calculated expression for the
transfer integral of the skyrmion we confirm existence of this highly
degenerate set .

It should be emphasized that the energy of the above addressed quantum small
$l$-skyrmions with bare radius $R< \lambda $  is always equal to the classical
single skyrmion energy independently both of $R$ and $l$ value. However, by
forming the appropriate combinations of single skyrmion states centered at
different space points with separation $R_{12}\geq 2\lambda$ the energy of the
classical state can be lowered. Addressing for example the combinations of two
different states with separation between them exceeding $2\lambda$ we obtain
usual bonding and antibonding states. Energy of such a bonding state is
$$
\frac{4 S^{2} J \pi}{1-\mid S_{12}(R)\mid}
$$
 and has a minimum at $R_{12}=2\lambda$. This implies that skyrmion states prefer
 to be
 separated a distance $2\lambda$ from each other  to lower the energy. Naturally,
this effect will lower the energy of the orbital skyrmions with rather large bare radius
  $R>\lambda$ as compared to the energy of classical skyrmion, but for large skyrmion radia    $\lambda$ this lowering is obviously negligible. Thus we may say that both large and small
$l$-skyrmions form unified zero Landau level.

The interaction of orbital skyrmions having
  the same orbital momentum but different bare radius
   will result in a complex skyrmionic energy spectrum, which analysis is a challenge
   for a further study. However, continual approximation used here is inadequate for calculation of that spectrum because real lattice on which skyrmion moves has no true
rotational symmetry.

 A further  elaboration of the quasiparticle approach to skyrmions
  implies a detailed description of
 the internal structure of the orbital skyrmions without any constraint on its
  bare radius, and realization of variational procedure $\cite{Tang}$. An appropriate work is
 in progress.

The research described in this publication was made possible in part by Award
No.REC-005 of the U.S. Civilian Research \& Development Foundation for the
Independent States of the Former Soviet Union (CRDF). The authors acknowledge a
partial support from the Russian Ministry of Education, grant \# 97-0-7.3-130.
We acknowledge stimulating discussions with G. Volovik and J. M. Tang.

\end{document}